\documentclass[10pt,aps,pra,showpacs,floatfix,amsmath,amssymb,raggedbottom,twocolumn]{revtex4-1}
\usepackage{amsmath,amssymb}
\usepackage{epsfig}
\usepackage{graphicx,latexsym}
\usepackage{mathrsfs}
\newcommand{\beq}{\begin{equation}}
\newcommand{\eeq}{\end{equation}}
\usepackage{color}
\usepackage{xcolor}
\usepackage{stix,amsmath}
\usepackage{tabularx}
\usepackage{multirow}

\begin{document}

\title{Manipulation of Gaussian derivative pulses and vector solitons in an anomalous-dispersion fiber laser}
\author{Wen-Hsuan Kuan}
\email{wenhsuan.kuan@gmail.com}
\author{Yu-Chia Leu}
\author{Kuei-Huei Lin}
\email{khlin@utaipei.edu.tw}
\affiliation{Department of Applied Physics and Chemistry, University of Taipei, Taiwan.}

\date{\today}

\begin{abstract}
Both positive- and negative-polarity Gaussian monocycle and doublet pulses, for which the pulse shapes are the first and second derivatives of Gaussian functions respectively, are generated in a ring-cavity erbium-doped fiber laser from polarization-locked vector solitons by using passive optical technology. The pulse states are switchable
and are found to be the superposition of bright and dark solitons with different widths, amplitudes, time delays, polarizations, and wavelengths. Qualitative analysis of the properties of vector solitons are performed by
solving coupled complex Ginzburg-Landau equations. By theoretical representing the envelopes of bright soliton by sech and dark soliton by tanh functions, the incoherent superposition of these two soliton components have simulated the experimental observations, and the underlying mechanisms on the formation for monocycle and doublet pulses are attributed to the polarization locking of bright and dark solitons. The results of tunable optical vectos solitons are compared with atomic solitons in the system of Bose-Einstein condensation. Since
governing equations for soliton generation in fiber lasers and Bose-Einstein condensation have many properties in common and thus the simulation and propagation of pulsating waves may open a new route to explore the classical solitary dynamics in nonlinear optics and its quantum analogy in ultralcold fields.
\end{abstract}


\maketitle

\section{Introduction}
Fiber lasers have many applications in laser science, optical communication, frequency metrology, {industrial application}, and biomedical technology, etc. Many rare-earth-doped fibers are used as the active media of fiber lasers. Among these active media, ytterbium-doped fibers and erbium-doped fibers have attracted much attention for either continuous-wave (CW) or pulsed fiber lasers \cite{Wise2008,Grelu2012,Lin2015,Nishizawa2014,Lin2015a,Kuan2018}, \textcolor{black}{since they can be widely used in laser machining and fiber-optic communications}. Semiconductor optical amplifiers have also been utilized as the gain media in fiber lasers to provide additional lasing wavelengths or functionalities \cite{Yang2004,Wang2015}. Various fiber lasers with different wavelengths and properties have become practical and useful test beds for laser physics and nonlinear dynamics. \textcolor{black}{Owing to the long interaction length in fiber waveguides, small cross section of the fiber cores, and the particular dispersion maps for
different cavity design, various nonlinear optical phenomena can be observed in fiber lasers.}

\textcolor{black}{The interests to study nonlinear dynamics of solitary waves may date back to 1960s since Zabusky and Kruskal studied the numerical solutions of the Korteveg-de Vries equation \cite{NailAkhmediev2008}. Subsequently, more soliton solutions were found in the systems of different fields described by, for example, the Burger equation \cite{Kutluay1999}, the sine-Gordon equation \cite{Rubinstein1970}, and the nonlinear Schr\"{o}dinger equation (NLSE).
In contrast to the extended-wave solutions, an ideal soliton or solitary wave will show the characteristic of the nonlinear eigen-modes, which connects the propagation speed of the solitary wave with its amplitude. A solitary wave shows the characteristic of neither dispersive nor dissipative propagation in the medium. A physical comparison shows that,
as the matter-wave soliton passing through an interface between two materials, the mass and the linear momentum have to be conserved, whereas it is the energy that has to be conserved in the scenario of an optical soliton.}
%
%
\textcolor{black}{Depending on the sign of fiber dispersion, either bright or dark solitons can be generated in passive single-mode fibers, which are governed by the NLSE \cite{Agrawal2013}. It was theoretically shown that the bright solitons can be formed in anomalous-dispersion fibers, while the dark solitons can be formed in normal-dispersion fibers. Regarded as a spatially or temporally coherent structure, the \textcolor{black}{ordinary} soliton solutions can exist provided that the nonlinearity can be counterbalanced by the dispersive property of the medium.}

\textcolor{black}{In the early days, the solitons or solitary waves were used to
describe the modes of nonlinear partial differential equations
that happened to be integrable by means of the inverse
scattering method.}
\textcolor{black}{However, the rigorous definition of soliton or solitary wave had been relaxed after 1990s as more and more observations had been reported in various systems. Although firstly proposed by Poincar\'{e} and Prigogine, the dissipative soliton concept has become a fundamental extension for solitons in the conservative and integrable systems. Whenever the effects of gain and loss are considered in a system, we have to turn to seek new stationary solutions beyond the ordinary solitons, which is concerned with the balance among gain, loss, dispersion, and nonlinearity.}
\textcolor{black}{It is worth of addressing that although gain and loss can co-exist in a laser system, in the steady-state the laser system would reach the status of maintaining the pulse shape in a particular position inside the laser cavity (self-consistency), and the net gain for a round-trip in the cavity becomes unity. Therefore, although the term 'dissipative solitons' are commonly used in the field of laser physics, it does not mean that the stored energy in the laser is decaying (dissipative). The dissipative solitons can be formed in a fiber laser of all-normal dispersion in combination with a positive nonlinear index, if there is in addition a spectral bandpass filtering effect and also an optical gain or amplification to compensate for the energy losses in the cavity.}
%

In birefringent fibers, a new type of optical solitons involving two orthogonal polarization components were found, which are termed as vector solitons \cite{Christodoulides1988}.
%
Owing to the strong optical intensity and large interaction length, several soliton families have been observed in fiber lasers. As compared with birefringent optical fibers, the formation of vector solitons in a fiber laser cavity is not only decided by the fiber birefringence, group velocity dispersion, and nonlinear Kerr effect, but also by the gain, loss, nonlinear absorption, optical filtering, and the cavity boundary condition. When operated in pulse states, most of the lasers produce bright pulses. However, it was shown that dark pulses or bright-dark pulses could be formed in fiber lasers \cite{Zhang2010,Zhang2009,Yuan2013,Fan2018}.
Shao et al. reported the formation of induced dark solitary pulses in a net anomalous dispersion cavity fiber laser, where the induced solitons were formed by the cross-phase modulation (XPM) between the two orthogonal polarization components of the fiber birefringence in the laser cavity \cite{Shao2015}. Zhang et al. observed an interesting type of vector dark soliton in a fiber ring laser, which consists of stable localized structures (polarization domain walls) separating the two orthogonal linear polarization eigenstates of the laser emission \cite{Zhang2010}. A series of vector solitons have been observed in weakly birefringent fiber laser cavities \cite{Zhang2009}. When polarization maintaining fibers are used in the cavity, a fiber laser is considered to be highly birefringent. Different from a weakly birefringent cavity, the group velocity mismatch between the two orthogonal polarization components cannot be ignored in a highly birefringent fiber laser \cite{Yuan2013}.

The observations and studies on dark and bright-dark solitons have enriched the research fields of nonlinear optics and fiber lasers. Mili\'{a}n et al. present experimental and numerical data on the supercontinuum generation in an optical fiber pumped in the normal dispersion range where the seeded dark and the spontaneously generated bright solitons contribute to the spectral broadening \cite{Milian2017}. Vector solitons can also been used in the alignment and rotation of individual plasmonic nanoparticles \cite{Tong2010}, vectorial control of matter magnetization \cite{Kanda2011}, and polarization division multiplexing in optical communication \cite{Chen2017}. Due to the potential applications in microwave photonics and \textcolor{black}{communication systems}, many optical methods have been proposed in past few years to generate Gaussian monocycle and doublet pulses, which have the pulse shapes of the first and second derivatives of Gaussian functions respectively \cite{Yao2009}. \textcolor{black}{The term "monocycle pulse" is commonly
used in the field of ultrawideband (UWB) communication.
It does not refer to a pulse of only one optical cycle. In addition, the
term "doublet pulse" do not refer to the soliton pair
consists of two bound solitons. The
distribution of UWB signals over optical fiber (UWBover-fiber)
is considered a promising solution for applications in broadband wireless communications.} However, most of the methods \textcolor{black}{used in generating
optical monocycle and doublet pulses} incorporated active technologies and multiple light sources, which are inevitably complicated and very expensive. The generation and manipulation of bright and dark solitons in fiber lasers may provide an economic solution to the optical pulse synthesizing
\textcolor{black}{of monocycle and doublet pulses}.

\textcolor{black}{From experimental perspectives}, in previous studies, we have achieved passive optical pulse generation in fiber lasers using a semiconductor optical amplifier as the gain medium \cite{Wang2012}. Various waveforms, including square wave, staircase wave, triangular wave, bright pulse, and dark pulse have been observed. With proper cavity tuning, it is likely to obtain more kinds of waveforms in fiber lasers.
In this paper, we first demonstrate the generation of Gaussian monocycle and doublet pulses in a ring-cavity erbium-doped fiber laser (EDFL) using passive optical technology. By adjusting intracavity polarization controllers, the fiber laser can be switched among continuous-wave, monocycle, and doublet states. The polarization of laser output are \textcolor{black}{also} measured to study the origin of pulse formation.
%

\textcolor{black}{From theoretical perspectives},
as only very few NLSEs can be solved analytically \cite{Ablowitz1991}, the pioneering work of Zakharov and Shabat \cite{Zakharov1972} stimulated great enthusiasm in finding the solutions of integrable NLSEs. Recently, Agalarov et. al. successfully modified the coupled NLSEs to Manakov and Makhankov U(n,m)-vector models and found vector solitons with unconventional dynamics for the passive fiber considering four-wave mixing terms \cite{Agalarov2015}.
However, in the low birefringent fiber with gain, loss, and optical filtering, the soliton dynamics can no longer be traced by solving NLSE with inverse scattering transform method.
\textcolor{black}{In recent years, Wise et al. verified that in real optical fibers, which are non-integrable and includes both conservative and non-conservative physics, their experimental observation of pulse-like solitary waves can be beautifully simulated by solving complex Ginzburg-Laudau Equation (CGLE) \cite{Wise2008}.} Inspired by Wise's study, in this work, we further employ qualitative analysis on the interactions of system parameters among group velocity dispersion (GVD), self-phase modulation (SPM), XPM, gain, loss, and optical filtering, as well as  their combined effects on the shaping of the bright and dark solitons, by solving coupled CGLEs with sech- and tanh-based trial functions.

\textcolor{black}{Since the systems of fiber lasers and Bose-Einstein condensation (BEC) have many things in common, the results of optical soliton generation are compared with atomic solitons in the system of BEC in the final part of this work. We start from reviewing the conservative systems in which the ordinary solitons can be generated \cite{Nistazakis2008,Yan2011,Becker2008,Khaykovich2002,Strecker2002,Cornish2006,Denschlag97}.
As the energy loss due to finite gain bandwidth in the optical fiber amplifier or the energy exchange between condensate and heat bath is considered, we can trace the dynamics of so-called dissipative solitons \cite{Achilleos2012}. Furthermore, when the solitons are found to
be generated in more complicated architectures with proper gain and loss that provide spatially- or temporally-dependent boundary conditions, the propagation of these solitons are likely to be drawn into periodic evolutions \cite{Rajendran2009,Arecchi2000}. In a birefringent fiber laser, the presence of XPM is crucial to the stable evolution and the generation of vector solition. Similarly, for matter-wave systems, the dark-bright solitons in spinor exciton-polariton BEC are promising to be created under nonresonant pumping \cite{Xu2018}.
While the domain wall solitons can be create in the two-component condensate with all-repulsive nonlinear interactions \cite{Coen2001,Kevrekidis2003}, we find that the optical domain wall between orthogonal polarizations can be manipulated by the adjustment of intracavity polarization in the fiber laser.}

\begin{center}
\begin{figure}[t!]
\includegraphics[width=0.3\textwidth]{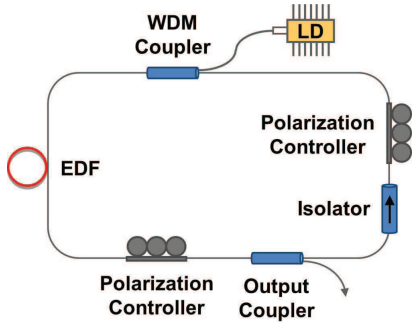}
\caption{(color online) Experimental setup of the ring-cavity erbium-doped fiber laser.}\label{Fig1}
\end{figure}
\end{center}

The paper is organized as follows. In section II, {we introduce the experimental setup}. In section III, {we discuss the experimental observations, as well as the details for the generation and manipulation of fiber solitons. Then, we construct the theoretical model to study the formation of vector solitons. We also compare the optical solitons with atomic solitons and decisively determine the mechanism in the formation of vector solitons in the anomalous-dispersion fibers.} Finally, in section IV we give a brief conclusion of the results.

\section{Experimental Setup}

The schematic setup of ring-cavity EDFL for Gaussian monocycle- and doublet-pulse generation is shown in Fig.~1. The cavity consists of a 980/1550 WDM coupler, a 90-cm erbium-doped fiber (LIEKKI Er80-4/125), two fiber polarization controllers, an output coupler, and a \textcolor{black}{polarization insensitive optical isolator}. The erbium-doped fiber is pumped by a laser diode (LD) with center wavelength of 974\,nm. \textcolor{black}{The cavity length is about 15.2\,m. In this EDFL, weak polarization dependence of bend loss is caused by
the polymer coating layer of the intra-cavity singlemode
fibers \cite{Wang2007}.}
The EDFL output signals are detected and characterized by a high-speed InGaAs detector, a digital oscilloscope, a sampling oscilloscope with O/E module, an optical power meter, an RF spectrum analyzer, and an optical spectrum analyzer. The polarization states of output laser pulses are analyzed by using a polarization controller and a polarization beam splitter.

\begin{center}
\begin{figure}[t!]
\includegraphics[width=0.45\textwidth]{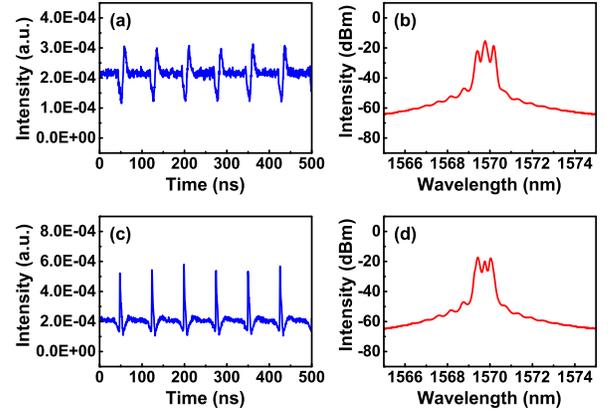}
\caption{(color online) Pulse trains and optical spectra of the monocycle pulses [(a)-(b)] and doublet pulses [(c)-(d)] generated in the ring-cavity EDFL.}\label{Fig2}
\end{figure}
\end{center}

\section{Results and Dsicussion}

\subsection{Generation and manipulation of fiber solitons}

For the ring-cavity EDFL of Fig.~1, the LD current is set to 210\,mA. After properly tuning of the intracavity polarization controllers (PCs), we obtain the monocycle pulses of dark-bright structure shown in Fig.~2(a), which has a repetition rate of 13.2\,MHz and average output power of 6.5\,mW. By further adjustments of the PCs, the relative delay and widths of the peak and valley can be tuned to obtain the doublet pulses of Fig.~2(c). The pulse repetition rate of doublets is the same as the monocycle pulses, but the output power is slightly decreased to 5.8\,mW. Another feature of different central hump intensities for two pulses can also be seen from the optical spectra of Figs.~2(b) and 2(d). For this ring-cavity EDFL, monocycle and doublet pulses of reversed shape (i.e., bright-dark for monocycle pulses and valley-centered for doublet pulses) have also been obtained by tuning the polarization controllers.

Figs.~3(a) and 3(b) show respectively the curve-fittings of monocycle and doublet pulses. The monocycle pulse can be fitted by the first derivative of Gaussian function as
\beq
y={{y}_{0}}+\left( \frac{A}{{{\mathrm{w} }^{2}}} \right)\,\left( x-{{x}_{c}} \right)\,\exp \left[ -\frac{{{(x-{{x}_{c}})}^{2}}}{2{{\mathrm{w} }^{2}}} \right],
\eeq
where $y_0 = {2.15}\times 10^{-4}$, $A = {5.98}\times10^{-4}$, $\mathrm{w} = {4.18}$\,ns, and $x_c = {54.8}$\,ns. The doublet pulse can be fitted by the second derivative of Gaussian function as
\beq
y={{y}_{0}}+\left( \frac{A}{{{\mathrm{w} }^{2}}} \right)\,\left[ 1-{{\left( x-{{x}_{c}} \right)}^{2}}/{{\mathrm{w} }^{2}} \right]\,\exp \left[ -\frac{{{(x-{{x}_{c}})}^{2}}}{2{{\mathrm{w}}^{2}}} \right],
\eeq
where $y_0 = {2.08}\times10^{-4}$, $A = {0.000184}\times10^{-4}$, $\mathrm{w} = {2.69}$\,ns, and $x_c = {48.6}$\,ns.

\begin{center}
\begin{figure}[t!]
\includegraphics[width=0.45\textwidth]{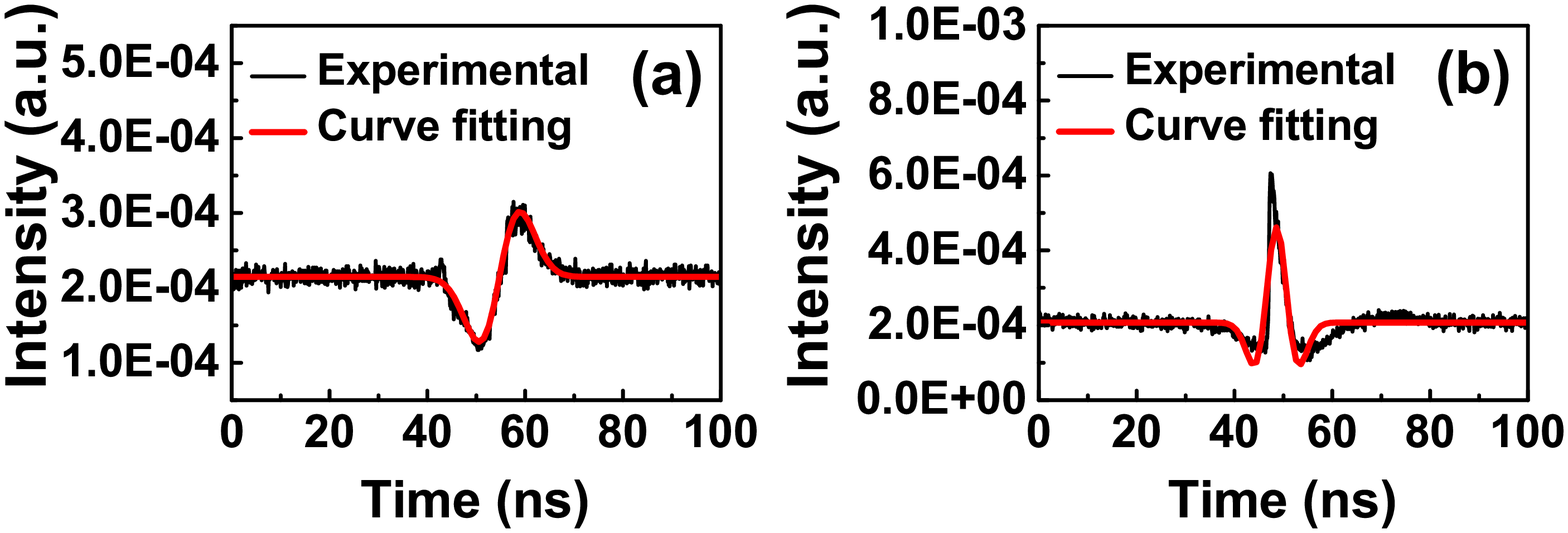}
\caption{(color online) Curve fittings of the (a) monocycle and (b) doublet pulses generated in the EDFL.}\label{Fig3}
\end{figure}
\end{center}

After analyzing the polarization states, we found that both the monocycle and doublet pulses are composed of two orthogonal polarization components. As shown in Fig.~4(a), the monocycle pulses are synthesized by bright (black curve) and dark solitons (red curve), which have the
characteristics of polarization domain wall solitons as in \cite{Zhang2010}. Owing to the incoherent
superposition of the two polarization components with fixed time delay and different structures,
the bright and dark solitons are locked to form the anti-symmetric monocycle pulse of Fig.~2(a).
The optical spectra of these two constituent polarization components are shown in Fig.~4(b), where
they are found to have different spectral distributions.

Fig.~4(c) shows the two polarization components of doublet pulses that are also synthesized by bright (black curve) and dark (red curve) solitons. We observed that the the extremities of
bright and dark components are almost coincident, and they become miscible without apparent domain walls. For doublets, the dark solitons have two spectral peaks, while the bright pulses have larger intensity toward longer wavelengths. Therefore, the underlying mechanism in the formation of monocycle and doublet pulses is similar, i.e., the polarization-locking of bright and dark solitons with different temporal widths, amplitudes, and time delays. When the centers of bright and dark pulses coincide, Gaussian doublet pulses are generated. On the other hand, when the centers of bright and dark pulses are detuned, Gaussian monocycle pulses will be formed. \textcolor{black}{Furthermore, the polarization of optical field within the monocycle pulses switches abruptly between two orthogonal polarizations to form a domain wall, whereas
the domain wall disappears for the doublet pulses since the polarization evolves gradually therein.}

\textcolor{black}{We observed that the time-bandwidth product (TBP) of
the soliton components in Fig.~3 and Fig.~4 is much larger
than the Fourier transform limit. }
\textcolor{black}{In the case of Fig.~3(b), the fitted width parameter of doublet is $w = 2.69$\,ns, for which the FWHM width of the bright component is 1.95\,ns [black curve in Fig.~4(c)] with spectral width of 0.13\,nm [black curve in Fig.~4(d)]. The pulsewidth is not transform-limited, because we do not introduce any dispersion compensation device in the cavity. In addition, the effective saturable absorption and the mode-locking strength induced by the weak polarization dependent loss in the laser cavity is not sufficient to lock all the cavity modes within the spectral width. In a previous
work, we have demonstrated that the generation
of nanosecond pulses could be caused by the
weakened intracavity pulse-shortening strength \cite{Kuan2018}. In fact, several research groups have reported the nanosecond mode-locked EDFLs. For example, Xia et al. demonstrated a mode-locked EDFL using a graphene-based saturable absorber. The pulse duration was measured to be 24\,ns, with center wavelength of 1569.5\,nm and 3-dB bandwidth of 1.2\,nm. The time-bandwidth product (TBP) of the generated nanosecond pulse is calculated to be 2806, which is much larger than the TBP value of 0.315 expected for transform-limited $\mathrm{sech}^2$ pulses \cite{Xia2014}. Xu et al. demonstrated a mode-locked nanosecond EDFL, in which the pulsewidth can be varied from 3\,ns to 20\,ns at a repetition rate between 1.54\,MHz and 200\,kHz by changing the cavity length \cite{Xu2012}.}

\begin{center}
\begin{figure}[t!]
\includegraphics[width=0.45\textwidth]{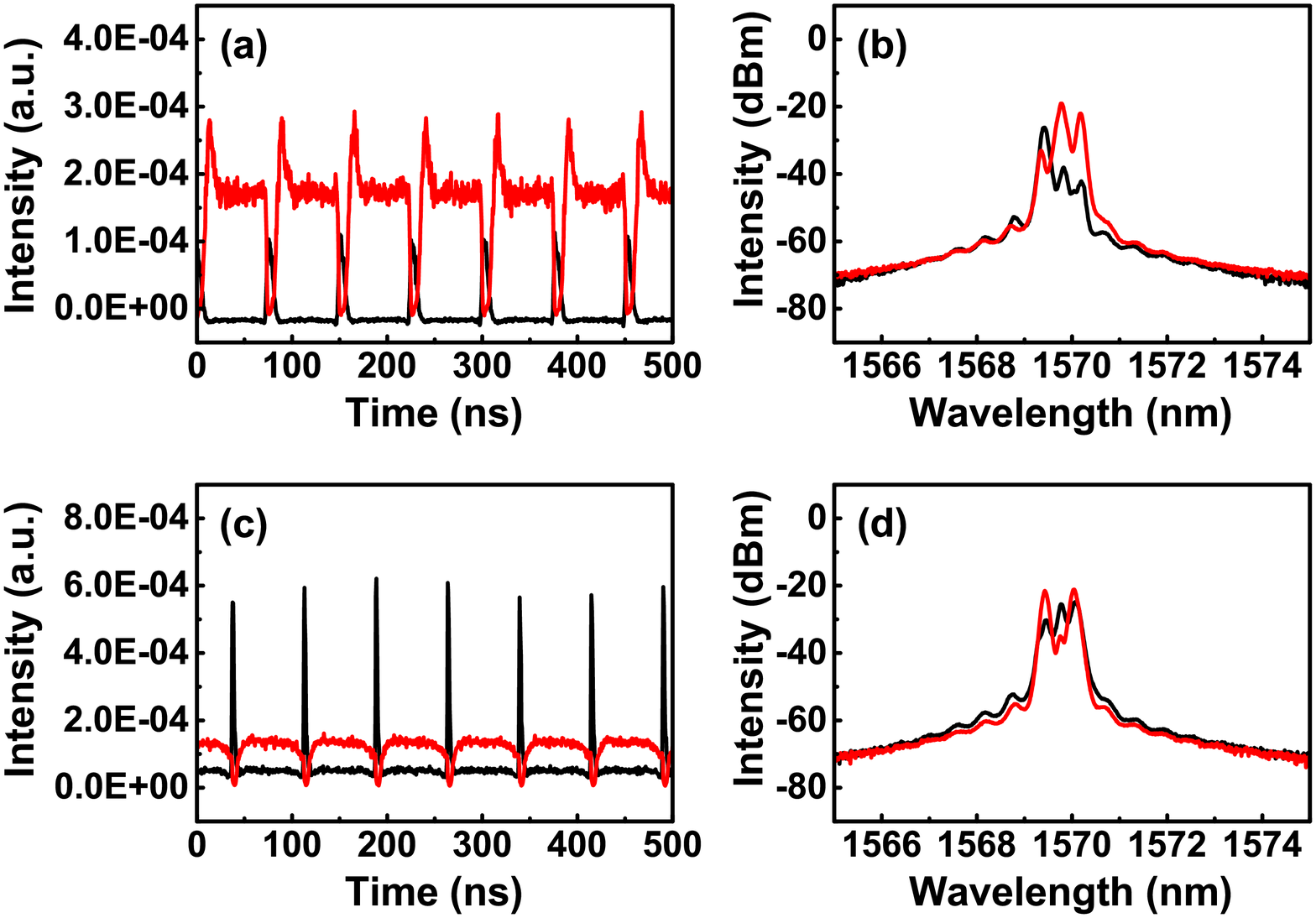}
\caption{(color online) Pulse trains and optical spectra of the polarization components in the monocycle pulses [(a)-(b)] and doublet pulses [(c)-(d)].}\label{Fig4}
\end{figure}
\end{center}

For this ring-cavity EDFL, Gaussian monocycle and doublet pulses with different polarities can also be generated. At LD pump current of 250\,mA, we properly adjusting the intracavity polarization controllers and have obtained the monocycle pulse of negative polarity (bright-dark) shown in Fig.~5(a), which has a reduced average output power of 4.3\,mW. The variation of output power can be attributed to the weak polarization-dependent loss induced by the bent intracavity fibers {\cite{Lin2014}}. The experimental data (black curve) can be fitted by the first derivative of Gaussian function plus a constant term (red curve). The monocycle pulses are found to be composed of bright and dark pulses, and the pulse train is shown in the
inset of Fig.~5(a). The separation between the peak and valley in the monocycle pulse is 9.3\,ns. We find that the positions, widths, and amplitudes of the bright and dark components can be tuned by adjusting the polarization controllers, and their polarization states are orthogonal. {Fig.~5(b)} is the optical spectrum of the bright-dark pulses, which has center wavelength of 1567.6\,nm and 3-dB bandwidth of 0.2\,nm. The RF spectrum is shown in the inset of Fig.~5(b). For this EDFL, monocycle pulse of positive
polarity (dark-bright) can also be obtained by carefully adjusting the polarization controllers, and the results are shown in Figs.~5(c) and 5(d). The separation between the valley and peak is 9.1\,ns, with optical spectrum centered at 1567.6\,nm and 3-dB bandwidth of 0.1\,nm. The output power is slightly increased to 4.4\,mW.

\begin{center}
\begin{figure}[t!]
\includegraphics[width=0.45\textwidth]{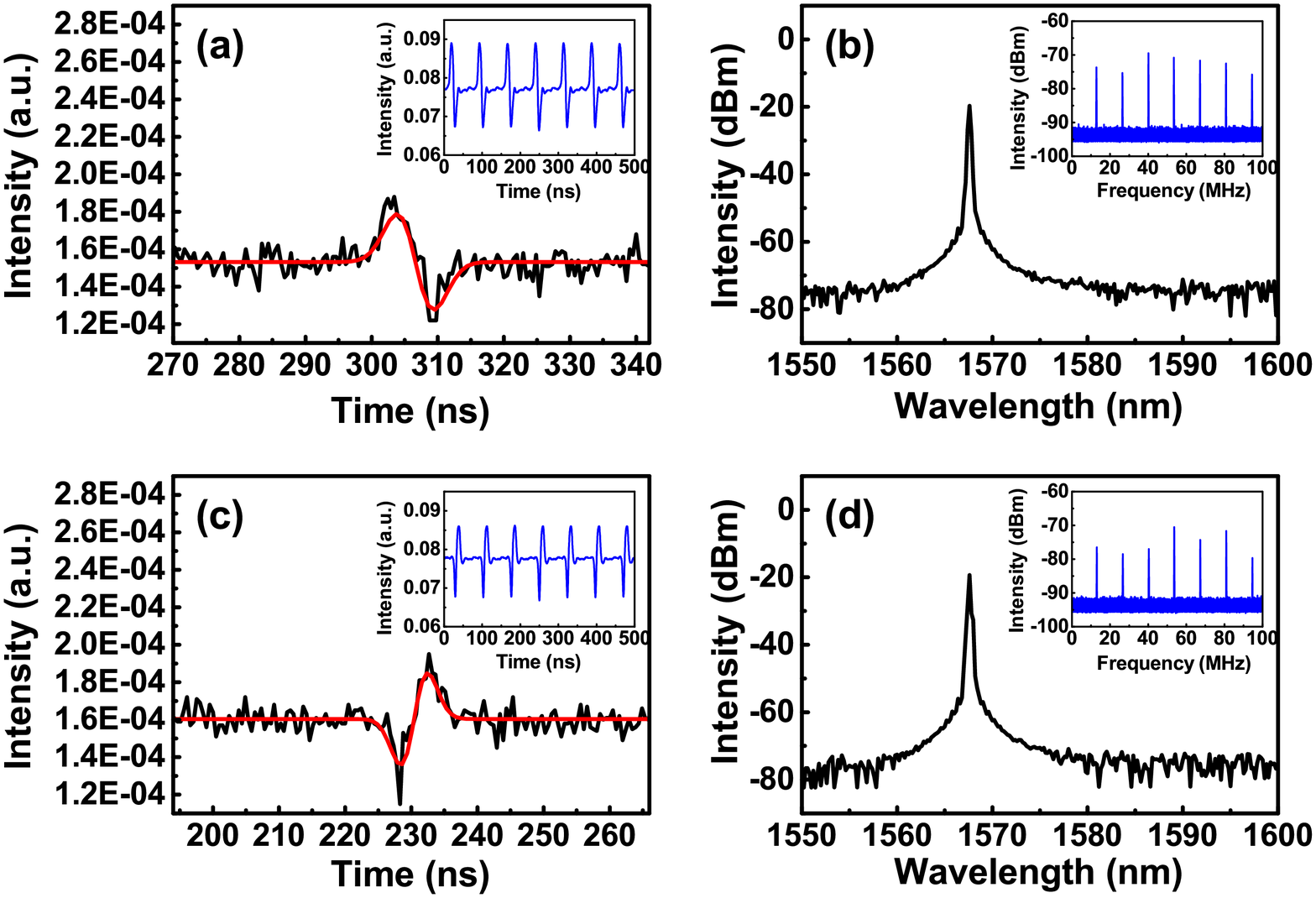}
\caption{(color online) (a) Temporal profile and (b) optical spectrum of the bright-dark pulses generated in the EDFL. (c) Temporal profile and (d) optical spectrum of the dark-bright pulses. The insets in (a) and (c) show the pulse trains, while the insets in (b) and (d) show the RF spectra.}\label{Fig5}
\end{figure}
\end{center}

By further adjustment of the intracavity polarization controllers, Gaussian doublet pulses of negative [Fig.~6(a)] and positive polarities [Fig.~6(c)] have been observed at the LD pump current of 250\,mA, and the output powers become 4.4\,mW and 4.9\,mW, respectively. Fig.~6(b) shows the optical spectrum of the negative-polarity doublet pulses, which has center wavelength of 1567.6\,nm and 3-dB bandwidth of 0.1\,nm. The center wavelength of positive-polarity doublet pulses in Fig.~6(d) is slightly shifted to 1568.4\,nm, with 3-dB bandwidth of 0.2\,nm.

\subsection{Theoretical modelling of vector solitons}

From the polarization analysis we claim that the two electric-field components of both monocycle and doublet pulses are mutually orthogonal. Furthermore, we also claim that the monocycle and doublet pulses can be synthesized from the incoherent superposition of bright and dark solitary waves. The propagation of vector solitons can be described by the coupled CGLEs
\begin{eqnarray}
   \frac{\partial u}{\partial z} &=&  i\beta u-\delta \frac{\partial u}{\partial t}-\frac{i{{k}^{''}}}{2}\frac{{{\partial }^{2}}u}{\partial {{t}^{2}}}+\frac{{{k}^{'''}}}{6}\frac{{{\partial }^{3}}u}{\partial {{t}^{3}}}
   + i\gamma \left( |u{{|}^{2}} \right.\nonumber\\
   && \left.+ {{\varepsilon }_{1}}|v{{|}^{2}} \right)u
    + i\gamma {{\epsilon }_{2}}{{v}^{2}}{{u}^{*}}{{e}^{-2i\rho z}}
    + \frac{g}{2}u+\frac{g}{2\Omega _{g}^{2}}\frac{{{\partial }^{2}}u}{\partial {{t}^{2}}} \\
  \frac{\partial v}{\partial z} &=& -i\beta v+\delta \frac{\partial v}{\partial t}-\frac{i{{k}^{''}}}{2}\frac{{{\partial }^{2}}v}{\partial {{t}^{2}}}+\frac{{{k}^{'''}}}{6}\frac{{{\partial }^{3}}v}{\partial {{t}^{3}}}
  + i\gamma \left( |v{{|}^{2}} \right. \nonumber\\
  && \left. +{{\varepsilon }_{1}}|u{{|}^{2}} \right)v
    + i\gamma {{\epsilon }_{2}}{{u}^{2}}{{v}^{*}}{{e}^{2i\rho z}}
  + \frac{g}{2}v+\frac{g}{2\Omega _{g}^{2}}\frac{{{\partial }^{2}}v}{\partial {{t}^{2}}},
\end{eqnarray}
in which $u$ and $v$ represent the envelopes of the two phase-locked polarization modes propagating along the erbium-doped fiber laser with net anomalous dispersion and birefringence, $u^*$ and $v^*$ are the complex conjugates of $u$ and $v$, $\rho = 2\beta$ is the wave number difference between two modes \textcolor{black}{and is associated with the birefringence of the fiber}, $2\delta$ is the negative of group velocity difference between two modes, $k''$ and $k'''$ are the $2^{nd}$ and $3^{rd}$ order dispersion coefficients, respectively, $\gamma = n_2\omega_0 / (c A_{eff})$, represents the nonlinearities in the fibers, where $n_2$ is the optical Kerr coefficient, $\omega_0$ is the central angular frequency and $A_{eff}$ is the effective mode area, $g$ is the saturable gain coefficient, and $\Omega_g$ is the bandwidth of the laser gain or \textcolor{black}{optical filtering effect}. Since the two orthogonal polarizations are locked together to form a vector soliton, they should have the same group velocity, and thus $\delta = 0$. Meanwhile, since we have pulses with durations much longer than tens of femtoseconds, the higher-order dispersion involving $k'''$ term can be reasonably ignored in the following calculations.
Taking the rotational invariance in an isotropic material, the contribution of the $3^{rd}$ order nonlinear polarization implies that $\epsilon_1 = 2/3$ and $\epsilon_2 = 1/3$, and the nonlinear term involving the birefringence would lead to the degenerate four-wave mixing. It's well known that in the highly birefringent fiber without net gain, the coupled CGLEs \textcolor{black}{with $\epsilon_1 = 1$ would recover Manakov's NLSEs and} can be solved by the inverse scattering method \cite{Manakov1974}. In this circumstance, the coupled CGLEs support the solutions for stable vector solitons of both fundamental mode and high-order modes.

\begin{center}
\begin{figure}[t!]
\includegraphics[width=0.45\textwidth]{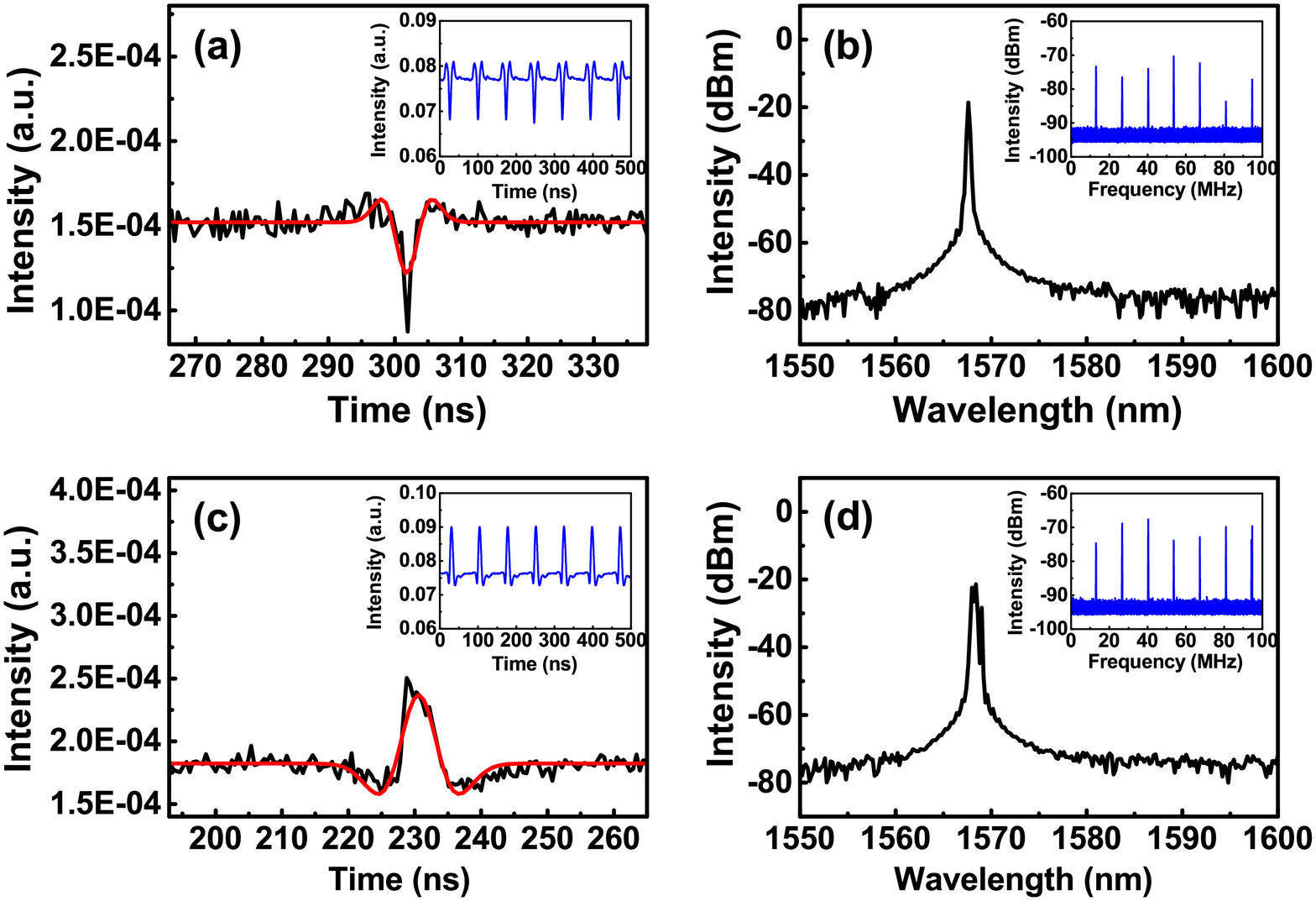}
\caption{(color online) (a) Temporal profile and (b) optical spectrum of the negative-polarity Gaussian doublet pulses generated in the EDFL. (c) Temporal profile and (d) optical spectrum of the positive-polarity pulses. The insets in (a) and (c) show the pulse trains, while the insets in (b) and (d) show the RF spectra.}\label{Fig6}
\end{figure}
\end{center}

\textcolor{black}{In reality}, the gain saturation and optical filtering (bandwidth-limiting) have to be considered in a real fiber laser system. {We assume that the fiber laser is composed of $N$-segments of fibers and each has length $L_n$, for $n = 1,\,2,\,3,\ldots,\, N$, with a specific set of parameters.} Instead of applying the inverse scattering method, a \textcolor{black}{sech-based} trial solution of the bright soliton {in one segment} is assumed to be in the form of
\begin{eqnarray}
u_n(z, t) &&= {u}_{0,n}\operatorname{sech}({{p}_{1,n}}(t-t_1))\exp\{-iq_n\ln [\cosh({{p}_{1,n}}t)]\}\nonumber\\
&& \times\exp[i(\sigma_n -{{a}_{1}}\rho_n )z], \label{eqn-usech}
\end{eqnarray}
and a dark solution has the \textcolor{black}{tanh-based} form of
\begin{eqnarray}
v_n(z, t) &&= v_{0,n}[\tanh({{p}_{2,n}}(t-t_2))-i\sqrt{1-{{B_n}^{2}}}]\nonumber\\
&&\times \exp\{-iq_n\ln [\cosh({{p}_{2,n}}t)] \}\exp[i(\sigma_n +{{a}_{2}}\rho_n )z], \label{eqn-vtanh}
\end{eqnarray}
{in which $t_1$ and $t_2$ are delay time. For a specific segment (for simplicity, we neglect the label $n$ in below)}, $u_0$ and $v_0$ denote the soliton amplitudes, the inverse of $p_1$ and $p_2$ denote the widths of the pulses, $q$ {sets the value of chirping}, and the parameter $B$ determines the intensity dip of the dark soliton. Only for $B = 1$ the intensity of the dip center would decrease to zero, otherwise for $|B| < 1$ the intensity of the dip center approaches a finite value and the dark soliton may be termed a grey soliton.

For weak birefringent fibers without considering gain, loss, optical filtering, chirping, and for solitons
with long temporal duration traveling therein, the substitutions of Eqs.~(\ref{eqn-usech}) and (\ref{eqn-vtanh}) into coupled CGLEs
{give ${{a}_{1}}={{a}_{2}}=\frac{1}{2}$, $q = 0$, $p_1^2 = p_2^2 = 8\beta/k''$, $u_0^2 = (\sigma - 6\beta)/\gamma$, and $v_0^2 = (\sigma + 2\beta)/\gamma$, where the introduction of a tunable parameter $\sigma$ sets the allowance for the formation of bright and dark solitons.
\textcolor{black}{The calculations exactly confirm Christodoulides's predictions \cite{Christodoulides1988} that in
a fiber with weak birefringence, the vector solitons can be established via XPM. In the region of normal dispersion, the bright soliton dominates the formation of the vector soliton, whereas it becomes the dark component to dominate the formation in the region of anomalous region.}

In real fibers without dropping gain, loss, and optical filtering effects, the solutions of
coupled CGLEs with trial functions of Eqs.~(\ref{eqn-usech}) and (\ref{eqn-vtanh}) bring out the same tuning parameters $a_1$ and $a_2$, and
\beq
q =\frac{3 k''\Omega _{g}^{2}}{2g}\pm \sqrt{\frac{9k'{{'}^{2}}\Omega _{g}^{4}}{4{{g}^{2}}}+2}, \label{eqn-qq}
\eeq
\beq
p_{1}^{2}=\frac{g}{2}{{\left[ \frac{g}{2\Omega _{g}^{2}}({{q}^{2}}-1)-k''q \right]}^{-1}}, \label{eqn-p1}
\eeq
\beq
p_{2}^{2}=\frac{g}{2}{{\left[ \frac{g}{\Omega _{g}^{2}}({{q}^{2}}-1)-\frac{3}{2}k''q \right]}^{-1}}, \label{eqn-p2}
\eeq
\beq
v_{0}^{2}=\frac{\sigma + 2\beta - {k''}p_{2}^{2}{{q}^{2}}/2}{\gamma (2-{{B}^{2}})}, \label{eqn-v02}
\eeq
\begin{eqnarray}
u_{0}^{2} &=& \frac{1}{\gamma }\left[ \sigma -2\beta +\frac{g}{2}q-{{p}_{1}}^{2}{k}''({{q}^{2}}+1)/2 \right]\nonumber\\
&& + \frac{(1-B)^2}{3}v_{0}^2. \label{eqn-u02}
\end{eqnarray}
\textcolor{black}{As stationary solitons can have
appreciable phase modulation, these calculations show that chirping is important for the solutions
of CGLE. The expression of Eq.~(7) implies that there are two possible solutions (bistability) for the chirping parameter $q$. For a fiber with anomalous dispersion, the vector solitons can be absolutely formed in the presence of up-chirping and negative birefringence. In this case, the bright soliton may dominate the formation. On the other hand, in a fiber of normal dispersion, the formation of vector solitons prefer down-chirping and positive birefringence. At the same time, the dark soliton turns to dominate the formation.}
\textcolor{black}{ Furthermore, as shown in Eqs.~(8) and (9), the
ratio between parameters $p_1$ and $p_2$ can be varied by
the gain coefficient $g$, leading to the possibility for
tuning the relative pulsewidths between bright and dark
solitons, as well as the generation of monocycle or doublet
pulses with reversed polarity.}
These expressions also reveal that with given system parameters, we are capable of qualitatively depicting the shape of solitons.
{Since the bright and the dark solitons have orthogonal polarizations, they form the fundamental solutions of the coupled CGLEs for each fiber segment. This implies that the
general solution of one segment should be written as the combination of two vectors, ie.,
$ \textbf{F}_n(z,t) = u_n(z,t)\hat{e}_u + v_n(z,t)\hat{e}_v $.}
{The magnitude of the coefficients $u_{0,n}$ and $v_{0,n}$ should be precisely determined with proper boundary conditions. In a ring-cavity, the boundary conditions require the conservation of pulse energy at the interface of two adjacent fibers and at the interface where the first fiber connects the last one.}

\begin{center}
\begin{figure}[t!]
\includegraphics[width=0.5\textwidth]{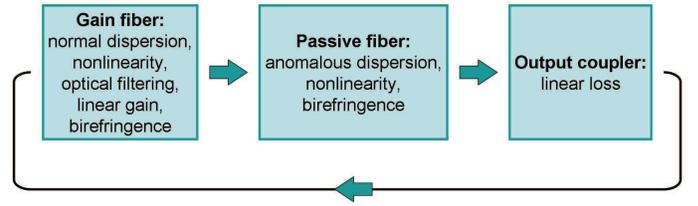}
\caption{(color online) The simplified fiber laser composed of a gain
fiber, a passive fiber, and an output coupler.} \label{Fig7}
\end{figure}
\end{center}

The simplest fiber laser with cavity length $L$ can be considered as composed of two fiber segments and \textcolor{black}{an output coupler, as shown in Fig.~\ref{Fig7}}. The
two fiber segments are the gain fiber and the passive fiber, with length $L_1$ and $L_2$, respectively. \textcolor{black}{The gain
fiber incorporates the effects of normal dispersion,
nonlinearity, optical filtering, linear gain, and birefringence,
while the passive fiber incorporates the effects of anomalous
dispersion, nonlinearity, and birefringence}. When the
solitons propagate inside the fiber of cross section $A_{eff,n}$,
the energy flowing across the interface of two fibers in a round-trip time $T$ is defined as
\beq
{E}_n = A_{eff,n} \int_0^T \left[|u_n(z,t)|^2 + (I_b - |v_n(z,t)|^2) \right]\,dt. \label{eqn-energycon}
\eeq
\textcolor{black}{in which $I_b$ is the CW background of the dark soliton, such as
those determined from the polarization-splitting measurements shown in Figs.~4(a) and 4(c).}
The criteria of energy conservation at the interfaces state that
\beq
\begin{array}{cc}
  {E}_1\mid_{z = L_1} = {E}_2\mid_{z = L_1}, \quad &  {E}_1\mid_{z = 0} = \alpha{E}_2\mid_{z = L}, \label{eqn-energybc}
\end{array}
\eeq
where $\alpha$ is the linear coupling loss coefficient of the output coupler.
{As the boundary conditions of Eq.~(\ref{eqn-energybc}) are employed, the magnitudes of $\sigma_1$ and $\sigma_2$ would no longer be free for tuning, but rather be completely decided.
As a consequence, with given sets of parameters, our derivations show that the shapes of the bright and the dark solitons are able to be clearly configured in the whole segments of the fiber laser.}
\textcolor{black}{Due to the weak polarization-dependent loss in a ring-cavity fiber laser \cite{Lin2014}, $g$ can be varied for different setting of the intracavity polarization controller. This may explain why the laser output power for 210-mA pumping is lower than that for 250-mA pumping in Sec.~IIIA.}

The adjustment of polarization controller will set the in-phase time [$t_1$ and $t_2$ in Eqs.~(5) and (6)] for the constituent frequencies of solitons, thus allows the tuning of peak locations of bright solitons (or the trough locations of the dark solitons). As a result, the relative delay of bright and dark components can be finely adjusted. The existence of fiber bending in the EDFL, including the three-paddle polarization controller, provides the weak polarization-dependent loss in the cavity. Therefore, the relative width and amplitude of the bright and dark solitons can also be adjusted. By adjusting the relative delay, width, and amplitude of the two polarization components, monocycle or doublet pulses with either positive or negative polarities are obtained.

The generation of Gaussian doublets in a ring-cavity EDFL using passive optical technology have not been reported previously. It is interesting to further study the origin of pulse formation for both the doublet and monocycle pulses. By analyzing the polarization states of the output pulses, we have found that the doublet pulses are composed of bright and dark pulses with different temporal width and orthogonal polarization states (Fig.~4), and the underlying mechanism in the formation of monocycle and doublet pulses is similar--the incoherent superposition of bright and dark pulses with different temporal widths, amplitudes, and time delays. When the centers of bright and dark pulses coincide, doublet pulses are generated. On the other hand, when the centers of bright and dark pulses are separated, monocycle pulses are formed. Fig.~\ref{Fig8} demonstrated this observation by theoretically superpose the intensities of a $\textrm{sech}^2$-shaped bright pulse and a $\textrm{tanh}^2$-shaped dark pulse. In Fig.~\ref{Fig8}(a), the dark (red dashed curve) and bright (black dashed curve) pulses have approximately equal widths and absolute amplitudes but their extremities are temporally staggered, which results in the dark-bright pulse (blue solid curve) similar to the experimentally observed monocycle pulses in Fig.~3(a). In Fig.~\ref{Fig8}(b), the extremities of a broad dark pulse with small absolute amplitude (red dashed curve) and a narrow bright pulse with large amplitude (black dashed curve) are temporally aligned, which results in the dark-bright-dark pulse (blue solid curve) similar to the experimentally observed doublet pulses in Fig.~3(b).

\begin{center}
\begin{figure}[t!]
\includegraphics[width=0.45\textwidth]{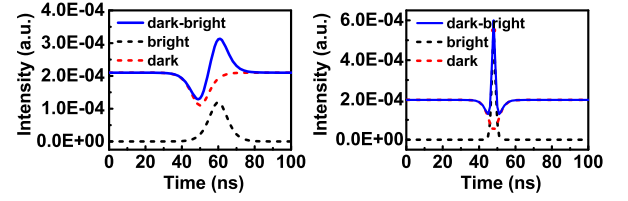}
\caption{(color online) Superpositions of dark and bright pulses to simulate the generation of (a) monocycle and (b) doublet pulses in Figs.~3(a) and 3(b), respectively.}\label{Fig8}
\end{figure}
\end{center}

\subsection{Optical soliton vs. matter-wave soliton}

The remarkable similarities between matter-wave solitons and optical
solitons reveal the close connection between atomic optics and light optics.
Their spatio-temporal similarities provide the opportunity to review the underlying properties and mechanisms from the opposite.
The analogies and the distinctions between matter-wave solitons and optical solitons are addressed below. We expect this investigation may pave the way for further generation of new-type solitons, as well as for property analysis in real fibers of the non-integrable system.

The dynamics of solitons in two-component Bose-Einstein condensates \textcolor{black}{with two different hyperfine states} can be described by the coupled Gross-Pitaevskii equations (GPEs),
\begin{align}
  & i\hbar \frac{\partial {{\psi }_{j}}(z,t)}{\partial t} = -\frac{{{\hbar }^{2}}}{2m}\frac{{{\partial }^{2}}{{\psi }_{j}}(z,t)}{\partial {{z}^{2}}}
   +{{V}_{ext}}(z) {\psi }_{j}(z,t) \nonumber\\
   &+\sum_{k=1}^2{{g}_{kj}}|{{\psi }_{k}}(z,t){{|}^{2}}{{\psi }_{j}}(z,t)
   -{{\mu }_{j}}{{\psi }_{j}}(z,t), \label{eqn-gpeu}
\end{align}
%
%
for $j = 1, 2$. In writing Eq.~(\ref{eqn-gpeu}) we have assumed that the transverse dynamics of cold atoms has been frozen out under extremely tight transverse trapping potentials so that theoretical studies of quasi-1D properties of the condensates and solitons within a shallow $V_{ext}$ can properly mimic the real experiments. The nonlinear terms represent the hard core collisions between two cold atoms taking places at ultralow temperatures, and the coefficients $g_{kj}$ count for the collision strength represented in terms of the s-wave scattering length. In the last terms of GPEs, $\mu_i$ represents the chemical potential of each component. The conservation of atom number provides the external constraint for Eq.~(\ref{eqn-gpeu}).

Different from the \textcolor{black}{simplest} optical NLSEs that mainly focus on addressing the pulse broadening caused either by normal dispersion ($ k'' > 0$) or by anomalous dispersion ($k'' < 0$) effect, and
the SPM and XPM due to the Kerr nonlinearity of the material, GPEs describe the Hamiltonian for the condensates.
Therefore for the condensates, the role of the kinetic energy terms drives the condensates as if they are moving with normal dispersions. The nonlinear coefficients $g_{kj}$ can be positive or negative, depending on whether the mutual interaction between two atoms is repulsive ($g_{kj} > 0$) or attractive ($g_{kj} < 0$). According to the naive balance condition, one could merely observe dark solitons within the condensate composing of repulsive atoms, and obtain bright solitons from the condensate composing of attractive atoms. In practice, via tuning scattering length near Feshbach resonances, it's possible to manipulate the interatomic collisions and realize magnetically or optically confined Bose-Einstein condensates as well as solitons in both regimes.
The development of density engineering and phase imprinting techniques bring us the possibility to create bright solitons, dark solitons, dark-bright solitons, and various soliton families in the one-component, multi-component, and spinor condensates {\cite{Pola2012,Nistazakis2008,Yan2011,Becker2008,Khaykovich2002,Strecker2002,Cornish2006}}.
It was reported that the presence of the spatial inhomogeneity strongly affect the motion, stability and interactions of the matter-wave solitons \cite{Busch2001}. As a signature of the soliton formation from condensates, an observation of undamped collective oscillations or the emission of non-dispersive solitary waves is expected \cite{Becker2008,Strecker2002}.

%

%

\textcolor{black}{Such as dissipative solitons can be generated in the optical fiber amplifier with finite gain bandwidth,} \textcolor{black}{the condensate solitons can also be supported in more complicated systems. While taking into account the finite-temperature effects, the interactions between condensates and thermal atoms emerge profoundly, and the nonequilibrium dynamics of BECs is used to be described by the stochastic GPE or dissipative GPE (DGPE),}
\textcolor{black}{
\begin{align}
   & (i-\gamma_j)\hbar \frac{\partial {{\psi }_{j}}(z,t)}{\partial t} = -\frac{{{\hbar }^{2}}}{2m}\frac{{{\partial }^{2}}{{\psi }_{j}}(z,t)}{\partial {{z}^{2}}}
   +{{V}_{ext}}(z) {\psi }_{j}(z,t) \nonumber\\
   &+\sum_{k=1}^2{{g}_{kj}}|{{\psi }_{k}}(z,t){{|}^{2}}{{\psi }_{j}}(z,t)
   -{{\mu }_{j}}{{\psi }_{j}}(z,t) + \eta_j(z,t),
\end{align}
}
\textcolor{black}{in which the thermally induced damping is simply represented in terms of a constant $\gamma$, and $\eta$ is a complex Gaussian function representing noise. However, due to the contact with reservoir, the solitons will eventually undergo the relaxation to the ground state of the system \cite{Achilleos2012}.  For two-component BEC, the evolution of the vector soliton demonstrates itself as a dissipative process, such
that, at sufficiently long times, the dark component will converge towards a Thomas-Fermi cloud  while the bright component will vanish.}

\textcolor{black}{The contact with the reservoir may also lead to certain dynamical equilibrium, if the condensate gets particle supplies therein.
Whenever the pumping rate is designed as a temporally periodic function, the revival and collapse
of the dark-bright solitons \cite{Rajendran2009} and the emission of atom laser have been observed \cite{Arecchi2000}.
In recent years, the microcavity exciton-polariton has emerged as a new type of BEC in two-dimensional semiconductors \cite{Deng2010,Kasprzak2006}. As the exciton-polariton condensate (EPC)
is intrinsically regarded as a nonresonantly pumped and nonequilibrium system, stable gap solitons in the optical lattices were found to be established \cite{Tanese2013,Cerda-Mendez2013, Cheng2018}. Just recently, the formation of long-lived dark-bright solitons in spinor polariton condensates is theoretically predicted \cite{Xu2018}. Similar to the adoption of CGLE for modelling optical solitons in the fiber laser system, the theoretical modelling of EPC incorporates spatial-temporal fluctuation-dissipation couplings between condensate and noncondensate particles, which can be implemented by applying the so-called complex GPE (CGPE) or generalized CGLE \cite{Arecchi2000},
\begin{align}
   & i\hbar \frac{\partial {{\psi }_{j}}(z,t)}{\partial t} = -\frac{{{\hbar }^{2}}}{2m}\frac{{{\partial }^{2}}{{\psi }_{j}}(z,t)}{\partial {{z}^{2}}}
   + i\hbar\frac{RD_r\Gamma}{2\gamma_u}\frac{{{\partial }^{2}}{{\psi }_{j}}(z,t)}{\partial {{z}^{2}}}\nonumber\\
   & + \left[\frac{i\hbar}{2}\left(\frac{R\Gamma}{\gamma_u}-\gamma_c\right)+{V}_{ext}(z)\right] {\psi }_{j}(z,t) \nonumber\\
   & + \left[\sum_{k=1}^2{{g}_{kj}}-\frac{iR\Gamma^2}{2\gamma_u^2}\right]|{{\psi }_{k}}(z,t){{|}^{2}}{{\psi }_{j}}(z,t)
   -{{\mu }_{j}}{{\psi }_{j}}(z,t). \label{eqn-CGLE}
\end{align}
In terms of the pumping rate function ($R$), the coupling rate between condensate and thermal atoms ($\Gamma$), the diffusion constant ($D_r$), the loss rate from the trap ($\gamma_u$), and the loss rate from the condensate ($\gamma_c$), Eq.~(\ref{eqn-CGLE}) demonstrates nonequilibrium coupled equations for an open system as the dissipative effects arising from diffusion from local fluctuation, linear gain and dissipation, and density dependent saturable loss are introduced into BECs. Whenever the diffusion term acts to enhance the dispersion effect and the presence of saturation gain lead to the nonlinear modulation, Eq.~(\ref{eqn-CGLE}) shows a parallel correspondence with the optical CGLE. A steady state solution can be found whenever the counterbalance among gain, loss, dispersion, and nonlinearities can be reached.
However, different from optical solitons, spatially-tight traps are indispensable in creating gap solitons. Moreover, these gap solitons
do not correspond to the ground state of the system. The observation with matter waves can only be achieved upon nonequilibrium atomic condensates.
Whenever the existence of bright (dark) soliton near the edge (center) of the Brillouin zones have been observed, the experimental verification of the spinor soliton generated from exciton-polariton BEC, which is dressed with right- and left-circularly polarized photons will provide
more confidence for the understanding of vector solitons in the matter wave systems.}

\begin{center}
\begin{figure}[b!]
\includegraphics[width=0.3\textwidth]{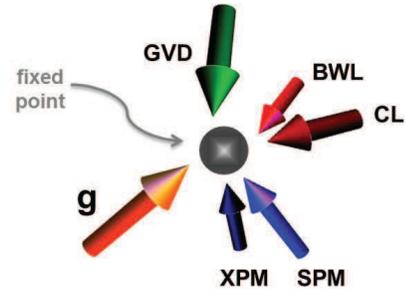}
\caption{(color online) Schematic diagram showing the generation of
a stable soliton by balancing various interactions in the fiber laser. }\label{Fig9}
\end{figure}
\end{center}
\begingroup
\squeezetable
\begin{table*}[t!]
\renewcommand{\arraystretch}{1.2}
\caption{Comparison between optical solitons and matter-wave solitons}
\label{table:1}
\begin{center}
\begin{tabular}{ l|l| l| l| l|}
\cline{2-5}
\cline{2-5}
& \multicolumn{2}{c}{Optical} & \multicolumn{2}{c|}{Atomic}\\ \cline{2-5}
& \multicolumn{1}{c|}{Physical example} &
    \multicolumn{1}{c|}{Governing equation} & \multicolumn{1}{|c|}{Physical example} & \multicolumn{1}{c|}{Governing equation} \\
\cline{1-5}
\cline{1-5}
\multicolumn{1}{|c|}{Ordinary}& \multicolumn{1}{c|}{Passive optical fiber} & \multicolumn{1}{c|}{} &\multicolumn{1}{|c|}{Conservative BEC }  & \multicolumn{1}{c|}{}\\
\multicolumn{1}{|c|}{soliton} & \multicolumn{1}{c|}{} & \multicolumn{1}{c|}{NLSE} &\multicolumn{1}{|c|}{}  & \multicolumn{1}{c|}{GPE}\\
\multicolumn{1}{|c|}{ }  & \multicolumn{1}{l|}{Ref: Phys. Rev. Lett. 45, 1095 (1980)} & \multicolumn{1}{c|}{} &\multicolumn{1}{|l|}{Ref: Science 287, 97 (2000) }  & \multicolumn{1}{c|}{}\\
\cline{1-5}
\multicolumn{1}{|c|}{Dissipative}& \multicolumn{1}{c|}{Optical fiber amplifier with finite gain bandwidth} & \multicolumn{1}{c|}{} &\multicolumn{1}{|c|}{Energy exchange between condensate}  & \multicolumn{1}{c|}{}\\
\multicolumn{1}{|c|}{soliton} & \multicolumn{1}{c|}{} & \multicolumn{1}{c|}{CGLE} &\multicolumn{1}{|c|}{and heat bath}  & \multicolumn{1}{c|}{{DGPE}}\\
\multicolumn{1}{|c|}{}  & \multicolumn{1}{l|}{Ref: Opt. Lett. 14, 943 (1989)} & \multicolumn{1}{c|}{} &\multicolumn{1}{|l|}{Ref: New J. Phys. 14, 055006 (2012)}  & \multicolumn{1}{c|}{}\\
\cline{1-5}
\multicolumn{1}{|c|}{Periodic }& \multicolumn{1}{c|}{Passively mode-locked fiber laser} & \multicolumn{1}{c|}{} &\multicolumn{1}{|c|}{Collapse-revival of soliton or atom laser}  & \multicolumn{1}{c|}{}\\
\multicolumn{1}{|c|}{soliton} & \multicolumn{1}{c|}{} & \multicolumn{1}{c|}{CGLE with $V_p(z)$} &\multicolumn{1}{|c|}{}  & \multicolumn{1}{c|}{CGPE with $V_p(t)$}\\
\multicolumn{1}{|c|}{evolution}  & \multicolumn{1}{l|}{Ref: Laser \& Photon. Rev. 2, 58 (2008)} & \multicolumn{1}{c|}{} &\multicolumn{1}{|l|}{Ref: J. Phys.
B.: At. Mol. And Opt. Phys. 42, 145307 (2009)}  & \multicolumn{1}{c|}{}\\
\cline{1-5}
\multicolumn{1}{|c|}{Vector }& \multicolumn{1}{c|}{Vector soliton generation } & \multicolumn{1}{c|}{} &\multicolumn{1}{|c|}{Vector soliton generation from exciton-polariton BEC}  & \multicolumn{1}{c|}{}\\
\multicolumn{1}{|c|}{soliton} & \multicolumn{1}{c|}{from fiber laser} & \multicolumn{1}{c|}{Coupled CGLE} &\multicolumn{1}{|c|}{}  & \multicolumn{1}{c|}{Coupled CGPE}\\
\multicolumn{1}{|c|}{}  & \multicolumn{1}{l|}{Ref: Opt. Express 21, 23866 (2013)} & \multicolumn{1}{c|}{} &\multicolumn{1}{|l|}{Ref: arXiv:1807.04401}  & \multicolumn{1}{c|}{}\\
\cline{1-5}
\end{tabular}
\end{center}
\end{table*}

\textcolor{black}{From aforementioned comparisons and discussions, finally we recognize that, just as a condensate is not equivalent to an atomic soliton, neither is a stimulated emission equivalent to an optical soliton. In any general system, whether a soliton can be stably created or not is highly depending on the counterbalance among various kinds of interactions of the system.
As a result, these effects will crucially determine whether a soliton or what kind of a soliton can be stably generated, and also dramatically influence the phenomena that can be observed.
Taking the fiber laser as an example system, the general concept of a soliton can be depicted in the schematic diagram of Fig.~\ref{Fig9}.}
In a lossless passive fiber, the formation of stable solitons are attributed to the balancing between SPM and GVD. To generate bright solitons, an intensity threshold should be reached, while the generation of dark solitons is thresholdless. However, in a real fiber laser system, various effects such as linear and saturable gains, SPM, XPM, bandwidth limiting (BWL), coupler loss (CL), and saturable loss should be considered. Similarly, in an open atomic system, the inclusion of the thermal contact with the reservoir leading to time-dependent gain, loss from condensate, and loss from trap are now known to have significant influence on the formation of atomic soliton and its life time. Moreover, such as the presence of XPM is crucial to the generation of vector solitons in a birefringent fiber laser, the sign and the strength of the nonlinear coupling between two-component condensates are deterministic to the matter-wave bound solitons. 
{Relative to an atomic dark-bright or bright-dark soltion that is formed in a superposition manner and displays a spatially localized structure,
the relative phase, pulsewidth, and the positions of amplitude
extremities for dark and bright solitons in the optical fibers can be easily manipulated via the adjustment of intracavity polarization.}
{While the domain wall solitons can be created in the two-component condensate with repulsive inter- and intra-atomic nonlinear interactions \cite{Coen2001,Kevrekidis2003}, we find that the domain wall between orthogonal optical polarizations can also be manipulated by the adjustment of intracavity polarization.}

\textcolor{black} {In the following we briefly summarize the comparison between optical and atomic solitons and list their features in Table~I. \\
(1) \textit{Ordinary soliton}---A conservative system provides the simplest scheme for the generation of solitons. In passive optical fibers, the generation of ordinary solitons can be described by the NLSE. For conservative BEC systems, the solitons can be generated by phase or density engineering and theoretically modelled by the GPE.\\
(2) \textit{Dissipative soliton}---For more realistic case, the energy loss due to finite gain bandwidth in the optical fiber amplifier or the energy exchange between condensate and heat bath is considered. In this situation, the dynamics of so-called dissipative solitons are described by the CGLE or DGPE for optical or atomic systems, respectively.\\
(3) \textit{Periodic soliton evolution}---When the solitons are found to be generated in the more complicated architectures, such as in the passively mode-locked fiber lasers that are sustained stably against linear and nonlinear losses by the external pumping, the propagation of these solitons are likely to be drawn into periodic evolutions. The formation of the optical solitons is governed by the CGLE, in which the loss together with the pumping serve as a spatially periodic potential to provide the boundary condition of energy continuity in each round trip.
For BECs in an open system allowing thermal contact with the reservoir in a regular manner, the emission of atomic soliton or atom lasers can be regarded as the parallel examples. The dynamics of the atomic soliton is described by the CGPE, in which the gain
is designed as a temporally periodic function to drive the periodic growth or collapse (emission)
of the condensate.\\
(4) \textit{Vector soliton }---The vector solitons with orthogonal polarizations can be generated
in the fiber lasers that have gain, loss, birefringence, and optical filtering effects. By solving coupled CGLEs, the dynamics of the vector solitons reveals that the presence of XPM is crucial to the stable evolution and the generation of vector solition.
For matter-wave systems, the dark-bright solitons in spinor exciton-polariton BEC can be created under nonresonant pumping. The dynamics of the vectorial gap solitons is described by the coupled CGPEs that include diffusion from local fluctuation, linear gain and dissipation, and density dependent saturable arising from the energy and particle exchange in the thermal contact with the reservoir. Different from the optical solitons, these gap solitons
do not correspond to the ground state of the system. Their observation with matter waves can only be achieved upon nonequilibrium atomic condensates.}

\section{Conclusion}

In this paper, Gaussian monocycle and doublet pulses of positive- and negative- polarities are generated in a ring-cavity erbium-doped fiber laser using passive optical technology. By adjusting polarization controllers, the fiber laser can be switched among continuous-wave, monocycle, and doublet states. After carefully examining the polarization of laser output, we find that both monocycle and doublet pulses are composed of bright and dark solitons having orthogonal polarizations. The temporal widths, amplitudes and time delays of the bright and dark components can be tuned by the intracavity PCs, and the underlying mechanisms on the formation of monocycle and doublet pulses are attributed to the polarization locking of bright and dark solitons. We have observed experimentally that a bright soliton and a dark soliton can be locked to form a monocycle bright-dark or dark-bright vector soliton, depending on the relative time delay between the two polarization components. It is also viable for a bright soliton to be embedded in a broader dark soliton to form a doublet vector soliton, or a dark soliton in a broader bright pulse to form a doublet of reversed polarity. Therefore, the polarization domains and shapes of the vector
solitons can be manipulated by the PC adjustments. Without resorting to expensive active components and complicated design, the method used here is simple for applications in optical pulse shaping and laser physics.

The formation of monocycle or doublet pulses are further demonstrated theoretically by the incoherent superposition of the bright and dark solitons with tanh envelope
functions, and the results shows good consistency with the experimental observations.
\textcolor{black}{The propagation of vector solitons in optical fibers can be
described by the coupled CGLEs, which is solved analytically
by assuming sech- and tanh-based solutions to obtain the
pulsewidths, chirping parameters, and soliton amplitudes.
For a given fiber, there exists two solutions (bistability) for
the pulse parameters, and the ratio between the pulsewidths
of bright and dark soliton components can be varied by the
gain coefficient, leading to the possibility of generating
monocycle or doublet pulses with reversed polarity.}
The results of tunable vector soliton generation are compared with matter-wave solitons in the system of BEC. We observed that for optical solitons, the polarization within the monocycle pulses switches abruptly between two orthogonal polarizations to form a domain wall, whereas
the domain wall disappears for the doublet pulses since the polarization evolves gradually therein.
Future studies can be conducted for the theoretical simulation of domain dynamics under different intracavity gain, dispersion, nonlinearity, and birefringence.
Since the general solutions for both CGLE and DGPE have not been found, the investigation and comparison may initiate some innovative ideas and pave the way for the generation of new type of solitons and the analysis of their properties in real fibers of the non-integrable system.

\section*{Acknowledgements}
This work is partially supported by the Ministry of Science and Technology, Taiwan, with grants MOST 107-2221-E-845-003 and MOST 104-2511-S-845-009-MY3.

\bibliographystyle{apsrev4-1}

%

\end{document}